\documentclass[letterpaper,12pt]{article}
\usepackage{amssymb, amsmath, amsfonts, natbib, graphicx} 
\usepackage{float, bm, appendix, multirow, setspace, footmisc, enumitem}
\usepackage[left=38mm,top=25mm,right=25mm,bottom=25mm]{geometry}
\DeclareMathOperator{\logit}{logit}
\newcommand{\M}{\mathcal{M}}

\newtheorem{theorem}{Theorem}

\newfloat{algorithm}{htbp}{loa}
\floatname{algorithm}{Algorithm}

\title{Semi-automatic selection of summary statistics for ABC model choice}

\author{{D}ennis {P}rangle\footnote{d.prangle@lancaster.ac.uk}{\ }\footnote{Department of Mathematics and Statistics, Lancaster University, UK}, {P}aul {F}earnhead\footnotemark[2], {M}urray {P} {C}ox\footnote{Allan Wilson Centre for Molecular Ecology and Evolution, Massey University, Palmerston North, New Zealand}{\ }\footnote{Institute of Fundamental Sciences, Massey University, Palmerston North, New Zealand}, {P}atrick {J} {B}iggs\footnotemark[3]{\ }\footnote{Infectious Disease Research Centre, Institute of Veterinary, Animal and Biomedical Sciences, Massey University, Palmerston North, New Zealand} \\ and {N}igel {P} {F}rench\footnotemark[3]{\ }\footnotemark[5]}

\date{\today}

\begin{document}

\maketitle

\begin{center}
{\bf Abstract}  
\end{center}

A central statistical goal is to choose between alternative explanatory models of data.  In many modern applications, such as population genetics, it is not possible to apply standard methods based on evaluating the likelihood functions of the models, as these are numerically intractable.  Approximate Bayesian computation (ABC) is a commonly used alternative for such situations.  ABC simulates data $x$ for many parameter values under each model, which is compared to the observed data $x_{\text{obs}}$.  More weight is placed on models under which $S(x)$ is close to $S(x_{\text{obs}})$, where $S$ maps data to a vector of summary statistics.  Previous work has shown the choice of $S$ is crucial to the efficiency and accuracy of ABC.  This paper provides a method to select good summary statistics for model choice.  It uses a preliminary step, simulating many $x$ values from all models and fitting regressions to this with the model as response.  The resulting model weight estimators are used as $S$ in an ABC analysis.  Theoretical results are given to justify this as approximating low dimensional sufficient statistics.  A substantive application is presented: choosing between competing coalescent models of demographic growth for \emph{Campylobacter jejuni} in New Zealand using multi-locus sequence typing data.


\section{Introduction}
The increasing availability of modern genetic data offers the possibility of learning more than ever before about the processes which generated it, for example the details of demographic change.  However, for stochastic models that incorporate a high level of detail, it is impractically costly to evaluate numerically the probability of a dataset, preventing inference by standard likelihood-based methods.  This has motivated the development of likelihood-free approaches, such as approximate Bayesian computation (ABC), which utilise the fact that simulating data from these models is relatively computationally cheap.

There is particular interest in using these methods to choose between explanatory models for observed data.  However \cite{Robert:2011} illustrated that applying ABC to model choice problems can produce highly inaccurate results.  This paper provides methods to address these concerns and improve the informativeness and efficiency of ABC model choice.  We focus on a particular application, inferring the demographic history of \emph{Campylobacter jejuni} in New Zealand from population genetic data.  This will be described in detail later.

A simple ABC algorithm operates by simulating data sets $x$ under various model and parameter pairs $(\M,\theta)$.  Pairs are accepted when $x$ is sufficiently close to the observed data $x_{\text{obs}}$.  This produces a sample of independent draws from an approximation to the Bayesian posterior distribution i.e.~that of $\M, \theta | x$.  Closeness is judged by the distance between vectors of \emph{summary statistics} $S(x_{\text{obs}})$ and $S(x)$.  Previous work (e.g.~\citealt{Blum:2010, Fearnhead:2012}) has shown that the quality of the approximations produced by ABC algorithms decays rapidly with the dimension of $S$.  This motivates finding low dimensional summary statistics.  However, it is crucial that these are also informative, as otherwise the problem of inaccurate results described by \cite{Robert:2011} can occur.

This paper sets out a method to choose $S(x)$ for use in model selection.  We give a theoretical result showing the existence of a low dimensional vector of statistics sufficient for model choice (under an appropriate definition given later).  Our method aims to estimate such a vector.  The idea is to use an extra simulation step to produce many $(\M, \theta, x)$ triples and then fit simple regression models of $\M$ on $x$.  Predictors from the fitted regressions form estimates of low dimensional sufficient statistics, and are used as $S$ in a main ABC analysis.  We refer to the approach as the \emph{semi-automatic method} as it adapts the method of the same name in \cite{Fearnhead:2012} which chooses $S$ by regressing $\theta$ on $x$ when the aim is inference of continuous parameters.

We expect that the targeted sufficient statistics are often complicated functions of the data which are hard to estimate globally.  To make the task easier, we advise that the regressions are based on data simulated, within each model, from a limited subset of parameter values which is judged by preliminary analysis to hold most of that model's posterior mass.  In other words, the simulation step mentioned above performs simulations from the models of interest following a truncation of their parameter supports.  The resulting $S$ can only be expected to perform well for choice between these truncated models.  A separate theoretical contribution of the paper is to relate results from such a choice to the original model choice problem.

Our approach of performing regressions based on simulated data is similar to \cite{Estoup:2012} who instead use linear discriminant analysis.  We expect our other contributions would also be useful to this approach.  Other work on ABC summary statistics has focused on validating a particular choice of $S$.  One approach is to run ABC analyses on a large number of simulated data sets to check whether $S$ provides accurate results \citep{Sousa:2011, Sjodin:2012}.  \cite{Marin:2012} give a complementary approach, identifying necessary and sufficient properties of $S$ for an ABC analysis to be consistent in an asymptotic regime corresponding to highly informative data.  Essentially, $S$ must have different asymptotic means under the models.  Given a choice of $S$, this property can be tested theoretically or through simulation.  Validation techniques are useful, but not sufficient, to choose $S$ for high dimensional genetic data where it is infeasible to compare all possible choices of $S$.  Our contribution is a method which can be applied in this setting to propose good choices of $S$.

Ideally the same ABC simulations would be used to provide inference on models and also their parameters.  The method we present provides summary statistics suitable for model choice only.  It would be desirable to augment them with informative summaries on model parameters, and we give an approach to do this that is specific to our main application.  General methods are an interesting topic for future research.

The remainder of the paper is organised as follows.  Section \ref{sec:background} describes ABC methods and our notation.  Section \ref{sec:theory} gives theoretical results on sufficiency, with proofs delayed until an appendix.  Section \ref{sec:method} explains our semi-automatic ABC method, and Section \ref{sec:examples} illustrates it for simple examples.  The application to \emph{Campylobacter} data is given in Section \ref{sec:appl}, and the article concludes with a discussion in Section \ref{sec:discussion}.  Further theoretical and simulation results are provided as supplementary material \citep{Prangle:2013}.

\section{Background} \label{sec:background}

Denote by $\M$ a random variable which can take values $\M_1, \M_2, \ldots, \M_M$, representing possible models.  Let $p_M$ be its prior mass function.  In an abuse of notation $\M$ will also denote a generic value of the variable, with usage clear from the context.  Each model represents a joint distribution $\pi(x, \theta | \M)$ on the data $x$ and parameters $\theta \in \Theta$.  This can be written as the product of prior and likelihood terms but we concentrate on the joint form for later convenience and to emphasise that the definition of a model includes a parameter prior.  Note that it is possible for the parameters under each model to belong to different spaces, in which case $\Theta$ is their union, and that $\theta$ will also be used to denote both a random variable and generic value.

Bayesian inference concentrates on $\pi(\theta | x, \M)$ -- the posterior distribution of parameters under a specific model -- and $\Pr(\M | x)$ -- the posterior model probabilities.  Inference on models can also be summarised using \emph{Bayes factors} $B_{ij} = \pi(x | \M_i) / \pi(x | \M_j)$; the ratio of the \emph{evidences} under models $\M_i$ and $\M_j$.  The Bayes factor does not involve $p_M$, but incorporating this information allows calculation of the ratio of posterior weights:
\[
\Pr(\M_i | x) / \Pr(\M_j | x) = B_{ij} p_M(\M_i) / p_M(\M_j).
\]

ABC is used in situations where it is possible to simulate $x | \M, \theta$ but evaluation of the density $\pi(x | \M, \theta)$ is impossible or impractically costly.  A simple approach to ABC inference is Algorithm \ref{alg:ABC_RS1} \citep{Grelaud:2009}.  

\begin{algorithm}[ht]
\rule{\textwidth}{0.7mm}
\begin{tabular}[h]{ll}
{\bf Input:} & Observed data $x_{\text{obs}}$, and a function $S(\cdot)$. \\
& A threshold $h \geq 0$ and a distance function $d(\cdot,\cdot)$. \\
& An integer $N>0$.\\
\\
{\bf Iterate:} & For $i=1,\ldots,N$
\end{tabular}
\begin{enumerate}[topsep=0pt]
\item Simulate $\M^*$ from $p_M(\M)$.
\item Simulate $\theta^*$ from $\pi(\theta | \M^*)$.
\item Simulate $x_\text{sim}$ from $\pi(x|\theta^*, \M^*)$.
\item Accept $(\M^*, \theta^*)$ if $d(S(x_{\text{obs}}), S(x_{\text{sim}})) \leq h$.
\end{enumerate}
\begin{tabular}[h]{ll}
{\bf Output:} & A set of accepted model and parameter pairs of the form $(\M^*, \theta^*)$.
\end{tabular}
\rule{\textwidth}{0.7mm}
\caption{Rejection sampling ABC incorporating model choice and parameter inference. \label{alg:ABC_RS1}}
\end{algorithm}
Letting $\mathbb{I}$ represent an indicator function, define
\begin{align*}
p_\text{ABC}(\M | S(x)) &\propto p_M(\M) \int \pi(S(x) | \M) \mathbb{I}[d(S(x_{\text{obs}}), S(x)) \leq h] dx, \\
\pi_\text{ABC}(\theta | \M, S(x)) &\propto
\pi(\theta) \int \pi(S(x) | \theta, \M) \mathbb{I}[d(S(x_{\text{obs}}), S(x)] \leq h) dx.
\end{align*}
Then the sample of $(\M, \theta)$ values output by Algorithm \ref{alg:ABC_RS1} is drawn from a distribution with conditionals $\pi_{\text{ABC}}(\theta | \M, S(x))$ and marginal $p_{\text{ABC}}(\M | S(x))$.

In the limit $h \to 0$, the ABC target distributions just defined converge on $\Pr(\M | S(x))$ and $\pi(\theta | \M, S(x))$.  However, reducing $h$ decreases the output sample size, increasing Monte Carlo approximation error.  A \emph{curse of dimensionality} result reviewed in the supplementary material shows that the rate of increase in error rises with the dimension of $S$.  This motivates a low dimensional $S$.  It is also important that $S$ is informative so that the limiting ABC targets approximate the posterior distributions $\Pr(\M | x)$ and $\pi(\theta | M, x)$ well.  Hence $S$ is a crucial tuning choice.

In practice, the results of Algorithm \ref{alg:ABC_RS1} can be highly variable if some prior model masses are small.  Algorithm \ref{alg:ABC_RS2} is a more stable alternative suggested by \cite{Grelaud:2009}.
\begin{algorithm}[ht]
\rule{\textwidth}{0.7mm}
\begin{enumerate}[topsep=0pt]
\item[] As Algorithm \ref{alg:ABC_RS1} except:
\item  Set $\M^*$ to $\M_1,\M_2,\ldots,\M_M$ with equal probability.
\end{enumerate}
\rule{\textwidth}{0.7mm}
\caption{A more stable modification of Algorithm \ref{alg:ABC_RS1}.}
\label{alg:ABC_RS2}
\end{algorithm}

Algorithm 2 samples $\M^*$ values from a uniform distribution rather than $p_M$, and it is necessary to correct the results to take this into account.  Let $n_i$ be the number of occurrences of $\M_i$ in the output sample.  Then $n_i / n_j$ is an estimator of the Bayes factor $B_{ij}$ and $n_i p_M(\M_i) / \sum_{j=1}^M n_j p_M(\M_j)$ is an estimator of $\Pr(\M_i | S(x))$.  The asymptotic and curse of dimensionality results outlined above continue to hold.  See \cite{Grelaud:2009} and the supplementary material for full details.

More efficient ABC model choice algorithms have been proposed, mainly based on sequential Monte Carlo (SMC) \cite[e.g.][]{Toni:2010, DelMoral:2011}.  However, the tuning issues just described remain.  The SMC algorithm of \cite{Toni:2010} is used later and described in the supplementary material.  Another approach to improve the quality of ABC results is to \emph{post-process} them.  This uses accepted parameters $\theta^{*,1}, \theta^{*,2}, \ldots$, models $\M^{*,1}, \M^{*,2}, \ldots$ and the corresponding simulations $x^{*,1}, x^{*,2}, \ldots$.  For parameter inference \emph{regression adjustment} \citep{Beaumont:2002, Blum/Francois:2010} fits a model $\theta=f(x,e)$, where $f$ is a deterministic function and $e$ a random residual, and outputs adjusted values $\theta'^{,i}=\hat{f}(x_{\text{obs}}, \hat{e}^i)$.  Model choice results can be post-processed by fitting a multinomial regression model $\Pr(\M|x) = g(x)$ and returning $\hat{g}(x_{\text{obs}})$ \citep{Beaumont:2008}.

\section{Theory} \label{sec:theory}

A statistic $S(x)$ of data $x$ is said to be \emph{Bayes sufficient} for parameter $\theta$ if $\theta | S(x)$ and $\theta | x$ have the same distribution for any prior distribution and almost all $x$ \citep{Kolmogorov:1942}.  This is a natural definition of sufficiency for ABC, as it shows that in an ideal ABC algorithm with $h \to 0$, the ABC target distribution equals the correct posterior when $S$ is used.  Throughout later sections of this paper we use ``sufficient'' to mean Bayes sufficient.
Theorem \ref{thm:suff} gives an alternative characterisation of Bayes sufficiency for $\M$ in the setting described in Section \ref{sec:background}.

\begin{theorem} \label{thm:suff}
Let $T(x)=\{ T_1(x), T_2(x), \ldots, T_{M-1}(x) \}$ where
\[
T_i(x) = \Pr(x | \M_i) / [\sum_{j=1}^M \Pr(x | \M_j)].
\]
Then $S$ is Bayes sufficient for $\M$ if and only if there exists a function $g$ such that $g[S(x)] = T(x)$ for almost all $x$.
\end{theorem}

Theorem \ref{thm:suff} shows that for any situation with $M$ models there are sufficient statistics for model choice of dimension $M-1$, namely the vector $T(x)$.  Furthermore, vectors $S(x)$ which can be transformed to $T(x)$ are also sufficient. 
\paragraph{Proof} See Appendix.

A sketch of the proof is as follows.  The theorem states that Bayes sufficiency of $S(x)$ for $\M$ is equivalent to there being a deterministic transformation from $S(x)$ to $T(x)$.  The latter vector is $M-1$ posterior probabilities given observations $x$ and uniform $p_M$.  Under uniform $p_M$, conditioning $\M$ on $S(x)$ satisfying this condition clearly recovers the posterior weights.  Reweighting can be used to show that the posterior is also recovered under any other $p_M$.  The converse can be shown by construction.

One particular sufficient choice of $S(x)$ used later is a vector of all Bayes factors under a one-to-one transformation.  Additionally, we note that a sufficient $S(x)$ may contain summaries which do not contribute to $T(x)$ but are useful for parameter inference.

Theorem \ref{thm:suff} is similar to Theorem 3a of \cite{Fearnhead:2012}, which shows that for continuous parameters $\theta$, $S(x) = E(\theta | x)$ is an optimal choice to estimate parameter means in terms of minimising quadratic error loss.  However this $S(x)$ is typically not sufficient for $\theta$.  Theorem \ref{thm:suff} is a stronger result for the case of model choice (or, equivalently, for estimating discrete parameters) showing the existence of low dimensional vectors of sufficient statistics.

\section{Method} \label{sec:method}

The low dimensional sufficient statistics described by Theorem \ref{thm:suff} are generally not available.  However their existence motivates an approach of approximating them from simulated data, and then using these approximations as $S(x)$ within ABC, as outlined in Algorithm \ref{alg:semiauto1}.  Step 2 requires some user input, as will be described in Section \ref{sec:fit}, so the method is referred to as ``semi-automatic ABC''.
\begin{algorithm}[h]
\rule{\textwidth}{0.7mm}
\begin{enumerate}[topsep=0pt]
  \item Simulate a large number of $(\M, \theta, x)$ triples.
  \item Calculate $S(x)$ by estimating sufficient statistics from simulations.
  \item Perform the ABC analysis using $S(x)$.
\end{enumerate}
\rule{\textwidth}{0.7mm}
\caption{Outline of simple semi-automatic ABC for model choice.  Full details of the steps are given in Sections \ref{sec:fit} and \ref{sec:method_other}.}
\label{alg:semiauto1}
\end{algorithm}

Sufficient statistics are likely to be highly complicated functions of the data due to the complexity of the models, and thus hard to approximate.  To make the task more tractable, we recommend some optional extra steps to give Algorithm \ref{alg:semiauto2}.  This simplifies the models by concentrating on the most likely parameter values.  We view this as replacing the models $\pi(\theta, x | \M_i)$ with \emph{truncated models}
\begin{equation} \label{eq:trunc}
\pi(\theta, x | \M_i') \propto \pi(\theta, x | \M_i) \mathbb{I}(\theta \in R_i),
\end{equation}
where $R_i$ is a \emph{training region} for model $\M_i$.  Calculation of $S$ is performed using data simulated from the truncated models.  The resulting $S$ estimates sufficient statistics for the choice between the truncated rather than original models.  Therefore the main ABC analysis must be performed between the truncated models, and, as will be shown in Section \ref{sec:method_other}, the results can be used to estimate the model choice posterior for the original problem.

\begin{algorithm}[h]
\rule{\textwidth}{0.7mm}
\begin{enumerate}[topsep=0pt]
  \item Perform an ABC \emph{pilot analysis} with ad-hoc summary statistics.  Use the output for each model to choose training regions $R_i$ of parameters which contain most of the posterior probability for each model $\M_i$.
  \item Simulate a large number of $(\M, \theta, x)$ triples using truncated models.
  \item Calculate $S(x)$ by estimating sufficient statistics from simulations.
  \item Perform the ABC \emph{main analysis} using $S(x)$ and truncated models.
  \item Use truncation correction to estimate posterior probabilities.
\end{enumerate}
\rule{\textwidth}{0.7mm}
\caption{Semi-automatic ABC for model choice with truncation steps.  Full details of the steps are given in Sections \ref{sec:fit} and \ref{sec:method_other}.}
\label{alg:semiauto2}
\end{algorithm}
The remainder of this section discusses the implementation of the steps in these algorithms in more detail.  Performance is assessed through simulation examples in Section \ref{sec:examples}.

\subsection{Calculating summary statistics} \label{sec:fit}

This section describes a logistic regression based approach to estimating sufficient statistics from simulated \emph{training data}.  A motivating example is the case of two models $\M_1$ and $\M_2$, with training data drawn from the joint distribution on $(\M,x)$, where $x = (x_1,x_2,\ldots,x_p)$.  Define $q(x)=\Pr(\M_1 | x)$.  This is clearly a sufficient statistic for $\M$.  Logistic regression can be used to fit
\begin{equation} \label{eq:logreg1}
\logit q(x) := \log\{q(x)/[1-q(x)]\} = \beta_0 + \sum_{i=1}^p \beta_i x_i.
\end{equation}
The fitted $\hat{q}(x)$ is an estimate of a sufficient statistic.  Note also that $q(x)/[1-q(x)]$ is the Bayes factor multiplied by a constant depending on the prior model weights.

To improve on the fit of \eqref{eq:logreg1} and cope with situations where $x$ is very high dimensional or not a fixed-length vector, in practice we fit instead
\begin{equation} \label{eq:logreg2}
\logit q(x) = \beta^T f(x),
\end{equation}
where $f(x)$ is a vector of transformation of $x$, including a constant term.  This can perform initial dimension reduction and introduce non-linear functions of the data into the regression.  Example choices of $f(\cdot)$ used later are 1) order statistics of raw data 2) a large number of summaries of genetic sequence data used in previous literature, and transformation of these (a constant term is also included in both cases).  To assist in the choice of $f(\cdot)$, regression diagnostics can be used, for example to compare the quality of the logistic regression fits for some $f_1(\cdot)$ and $f_2(\cdot)$.  The supplementary material gives examples in which cross-validation estimates of the deviance are compared.

In general the aim is to calculate $S$ for choice between models $\M_1, \M_2, \ldots, \M_M$, which for this discussion may represent original or truncated models.  Fix a pair of distinct models, $\M_i$ and $\M_j$, and consider the subset of training data made up of only the simulations from these models.  Logistic regression can be used as above to estimate the probability $q_{ij}$ of $\M_i|x$ under the $(\M,x)$ distribution for this training data subset.  This is repeated for each pair of distinct models, and results in a vector of one-to-one transformations of Bayes factors.  This target was shown to be sufficient for $\M$ in Section \ref{sec:theory}.



The logistic regression method set out above gives $\dim(S)=M(M-1)/2$, whereas Theorem \ref{thm:suff} shows there are sufficient statistics of dimension $M-1$.  Alternative regression methods can be used to give $\dim(S)=M-1$, for example estimating an appropriate subset of the Bayes factors or multinomial regression. 
In this paper we consider only examples with $M \leq 3$ so the logistic regression approach has limited excess dimension.  We believe it also aids robustness. Even if the logistic regression for one pair of models fits poorly (as is the case in the \emph{Campylobacter} application), the others can still allow a good overall estimate of sufficient statistics.  

\subsection{Other steps} \label{sec:method_other}

\paragraph{Pilot analysis}

The pilot ABC analysis uses an ad-hoc choice of summary statistics $S_{\text{pilot}}$.  The purpose of the pilot analysis is to roughly identify regions containing most of the posterior mass, so the procedure should be reasonably robust to the choice of $S_{\text{pilot}}$.  \cite{Fearnhead:2012} illustrate this argument by example.  Validation tests could also be performed to test the quality of ABC output from analysing simulated data using $S_{\text{pilot}}$.

In our implementation the pilot uses an ABC model choice algorithm such as Algorithm \ref{alg:ABC_RS2}.  An alternative approach would be to perform a separate pilot run for each model, focusing only on finding training regions, rather than initial model choice analysis.  We did not investigate this as a pilot analysis incorporating model choice has useful properties.  The estimated posterior can serve as a verification that the final results appear sensible.  Also, if the pilot results are sufficiently convincing in showing that certain models are incompatible with the data, they could be ruled out at this stage saving computational resources.  

\paragraph{Training region choice}

The training region $R_i$ for model $\M_i'$ should cover most of the posterior mass.  Our implementation is to choose a hypercube, with the range of each parameter being the interval of sampled values. 

\paragraph{Simulating data}

We generate training data from the distribution on $(\M, \theta, x)$ defined by the priors and models (or truncated models).  An alternative model distribution can be used without affecting the arguments in Section \ref{sec:fit} showing that the fitted summary statistics are estimates of sufficient statistics.  This would be useful if some prior model weights are too small to fit all regressions well.



\paragraph{Truncation correction}

Results of the main ABC analysis choosing between truncated models can be used to estimate those for the original model choice problem by the following consequence of \eqref{eq:trunc}:
\[
\pi(x | \M_i) = r_i \pi(x | \M_i'), 
\quad \text{where } r_i = \Pr(\theta \in R_i | \M_i) / \Pr(\theta \in R_i | x, \M_i).
\]
That is, the evidence of $\M_i$ equals that of $\M_i'$ multiplied by $r_i$, the ratio of the prior and posterior probabilities of $R_i$.  This allows estimation of Bayes factors or posterior probabilities for the original models given $r_i$ values.  As $R_i$ is chosen with the aim of containing most of the posterior mass, we estimate its posterior probability by 1, giving an estimate $\hat{r}_i = \Pr(\theta \in R_i | \M_i)$.  This prior probability can usually be calculated directly when $R_i$ is a hypercube.

\section{Examples} \label{sec:examples}

To illustrate our semi-automatic ABC method, we apply it to three simple binary model selection examples from the literature \citep{Didelot:2011, Marin:2012}, and extend one of these to a 3 model example.  The examples are summarised in Table \ref{tab:examples}.  The binary examples are the first two models in each letter group, and the 3 model example is the full A group.  In each case the data are 100 independent draws from one of the models and the models have equal prior probabilities.  All ABC analyses were performed using Algorithm \ref{alg:ABC_RS2}.

\begin{table}[htp]
\begin{center}
\begin{tabular}{c|ll}
Name & Model & Prior \\
\hline
A1 & Poisson($\theta$) & $\theta \sim \text{Exponential}(1)$ \\
A2 & Geometric($\theta$) & $\theta \sim \text{Uniform}(0,1)$ \\
A3 & Binomial($10, \theta$) & $\theta \sim \text{Beta}(1,9)$ \\
B1 & Laplace($\theta, 1/\sqrt{2}$) & $\theta \sim \text{Normal}(0,2^2)$ \\
B2 & Normal($\theta, 1$) & $\theta \sim \text{Normal}(0,2^2)$ \\
C1 & gk($0,1,0,k$) & $k \sim \text{Unif}(-0.5,5)$ \\
C2 & gk($0,1,g,k$) & $(g,k) \sim \text{Unif}([0,4] \times [-0.5,5])$
\end{tabular}
\caption{Models used in the examples of Section \ref{sec:examples}.  For details of the $g$-and-$k$ distribution see \cite{Rayner:2002}.}
\label{tab:examples}
\end{center}
\end{table}

\paragraph{Binary model selection}

The semi-automatic ABC method of Algorithm \ref{alg:semiauto2} was implemented starting with a pilot analysis using $S_{10}(x)=(x^{(5)}, x^{(15)}, \ldots, x^{(95)})$ where $x^{(i)}$ is the $i$th order statistic.  Model choice summary statistics were fitted as described in Section \ref{sec:fit} using $f(x)=(1,x^{(1)},x^{(2)},\ldots,x^{(100)})$.  No other summaries were added for parameter inference.  The analysis used $2 \times 10^4$ simulations, one quarter for the pilot and the rest used for both summary statistic fitting and the main analysis.  The pilot and main analysis both accepted 100 simulations.  Some alternative ABC analyses on the data were performed, each using the same total number of simulations and acceptances.  Firstly, the analysis was repeated using Algorithm \ref{alg:semiauto1}.  Secondly, standard ABC analyses were performed with Algorithm \ref{alg:ABC_RS2} using (a) $S=S_{10}$ (b) $S$ as in \cite{Marin:2012}; 4th and 6th moments for B, 10\% and 90\% quantiles for C.  All ABC analyses used the following distance
\begin{equation} \label{eq:scaled dist}
d(x,y) = \left[ \sum_{i=1}^p (x_i-y_i)^2 / \hat{\sigma}_i^2  \right]^{1/2},
\end{equation}
i.e.~Euclidean distance between $p$-dimensional summary statistics normalised by estimated standard deviations, $\hat{\sigma}_i$.  The latter were estimated from the simulated data.

\begin{figure}[htp] \begin{center}
  \includegraphics[totalheight=4in]{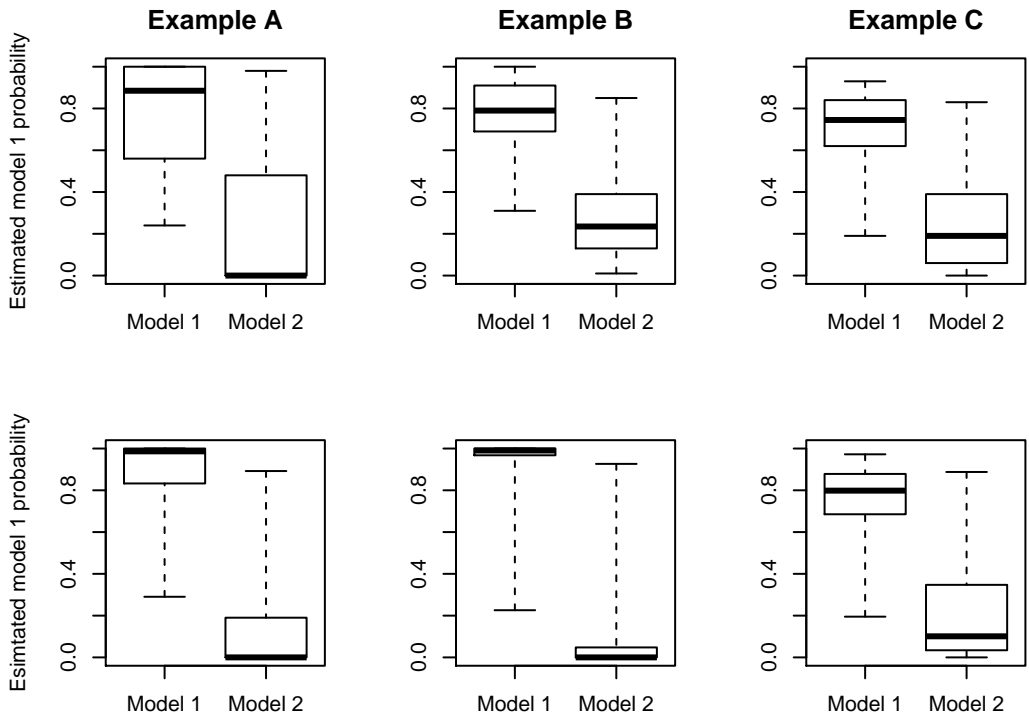}
  \caption{Boxplots of posterior probabilities of model 1 estimated by ABC (without post-processing) for 100 simulated datasets in each of three binary model comparison examples.  The boxplots show quartiles of the values.  Within each graph results are split by which model generated the data.  The top row uses $S=S_{10}$, and the second row chooses $S$ by semi-automatic ABC (Algorithm \ref{alg:semiauto2}).  The columns represent three model choice examples detailed in Table \ref{tab:examples}.} \label{fig:examples}
\end{center} \end{figure}

Figure \ref{fig:examples} shows estimated posterior probabilities for $S_{10}$ and Algorithm \ref{alg:semiauto2}.  Numerical summaries of estimation quality are given in Table \ref{tab:exres}.  This reports the entropic loss \citep{Robert:1996},
\[
-\sum_{i=1}^{100} \log \hat{\Pr}(m_{0,i} | x_{\text{obs},i}),
\]
the negative log probability of the correct models $m_{0,1}, \ldots, m_{0,100}$ estimated from the corresponding simulated datasets $x_{\text{obs},1}, \ldots, x_{\text{obs},100}$.  Also reported is the misallocation rate; the proportion of datasets where the highest weighted model was not the correct model.  Our method provides an improvement in all scenarios, although this is modest for example C.  The use of the truncation steps from Algorithm \ref{alg:semiauto2} is shown to sometimes be crucial; when Algorithm \ref{alg:semiauto1}, which omits these, is used instead, the results for example C are the worst of all methods.  However the effect is problem dependent; in example B it made little difference.  Exact posterior calculations are possible for examples A and B (the required Laplace marginal likelihood calculations are described in Appendix 1 from version 1 of \citealt{Marin:2012}), and in both cases Algorithm \ref{alg:semiauto2} provides comparable results.

We attempted to apply post-processing by the method of \cite{Beaumont:2008}.  For example A this was usually not possible as there was no variation in the accepted summaries, which were discrete in this case, or because all acceptances were for a single model.  For the other examples, it had little effect on entropic loss or misallocation rate, so these are not reported.  

\begin{table}[htp]
\begin{center}
\begin{tabular}{c|llll}
& \multicolumn{4}{c}{Example} \\
\multicolumn{1}{c|}{Summary statistics} & \multicolumn{1}{c}{Binary A} & \multicolumn{1}{c}{Binary B} & \multicolumn{1}{c}{Binary C} & \multicolumn{1}{c}{3 models} \\
\hline
$S_{10}$ & 33.0 (17\%) & 33.5 (11\%) & 43.0 (16\%) & 70.7 (39\%) \\
From literature & \multicolumn{1}{c}{-} & 55.3 (25\%) & 40.9 (20\%) & \multicolumn{1}{c}{-} \\
From Algorithm \ref{alg:semiauto1} & 30.2 (14\%) & 13.5 (5\%) & $\infty$ (21\%) & \multicolumn{1}{c}{65.9 (42\%)} \\
From Algorithm \ref{alg:semiauto2} & 19.8 (15\%) & 13.9 (7\%) & 38.4 (14\%) & 58.9 (33\%) \\
\hline
Posterior & 19.8 (12\%) & 15.6 (8\%) & \multicolumn{1}{c}{-} & 58.1 (36\%)
\end{tabular}
\caption{Entropic loss and misallocation rate (in brackets) from several ABC analyses of 100 simulated datasets in each of four model comparison examples, detailed in Table \ref{tab:examples}.  The final row shows values under the exact posterior, where these are available, for comparison.}
\label{tab:exres}
\end{center}
\end{table}

\begin{figure}[htp] \begin{center}
  \includegraphics[totalheight=4in]{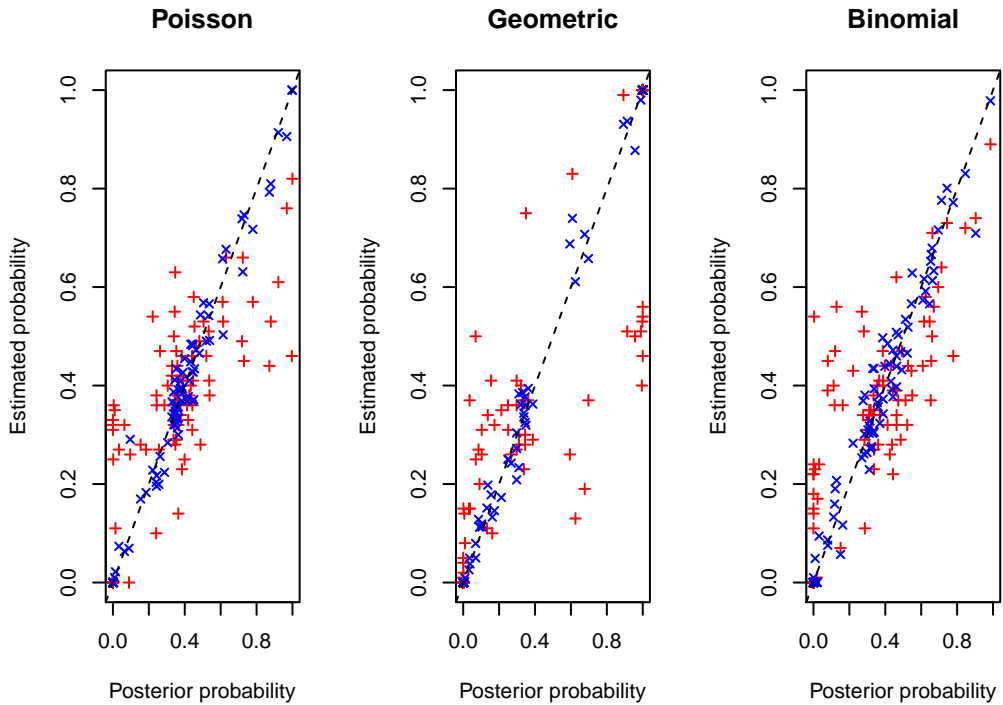}
  \caption{Plots of true posterior model weight against ABC estimates for 100 simulated datasets in a three model example.  Pluses are for $S=S_{10}$ and crosses for $S$ chosen by semi-automatic ABC (Algorithm \ref{alg:semiauto2}).}
  \label{fig:PGB}
\end{center} \end{figure}

\paragraph{Selection between three models}
Algorithms \ref{alg:semiauto1} and \ref{alg:semiauto2} were implemented as for the two model examples, with the addition that three summary statistics were fitted, corresponding to three pairs of models.  Figure \ref{fig:PGB} plots exact posterior probabilities against ABC estimates, and shows that Algorithm \ref{alg:semiauto2} performs better than the comparison analysis using $S=S_{10}$.  This is confirmed by the quantitative summaries in Table \ref{tab:exres}, which also shows that Algorithm \ref{alg:semiauto2} outperforms Algorithm \ref{alg:semiauto1} and achieves comparable results to the true posterior values.  Post-processing results are not shown because, as mentioned above, they could usually not be calculated for this example.

\section[Application]{Application} \label{sec:appl}

\emph{Campylobacter jejuni} and \emph{C.~coli} are bacterial pathogens that are a major cause of human gastroenteritis around the world \citep{Humphrey:2007}.  They are considered commensals of a wide variety of animals, including poultry, ruminants and wild birds, and human infection occurs as a result of ingesting contaminated food or drinking water and via direct contact with animal faeces \citep{Savill:2003}.  New Zealand has very high rates of campylobacteriosis and an investigation into the source of human infection \citep{Mullner:2009} has generated a large dataset of isolates from humans and animals that have been characterized by multilocus sequence typing, MLST \citep{Dingle:2001}.  The dataset of \emph{C.~jejuni} and \emph{C.~coli} isolates from New Zealand has been used to inform control policy \citep{Sears:2011} and to estimate evolutionary parameters, such as the rates of mutation and recombination \citep{Yu:2012}.  We focus on the question of demographic history, which is of particular interest in New Zealand due to the relatively recent colonization by man and the unique pattern of animal introductions (both wildlife and domestic animals) \citep{Atkinson:1993}. We ask: can we detect historic growth in the effective population size, and if so, does it correspond to a particular historical event? The relative isolation of this location means that neglecting ongoing exchange with outside populations is reasonably realistic. MLST data are available for over 3000 isolates from a variety of hosts.  

We present our methods and results below, with a discussion given in Section \ref{sec:campy discussion}.  Some further details are provided as supplementary material.

\subsection[Models and priors]{Models and priors}

We modelled the \emph{C.~jejuni} data using a coalescent model using the Jukes-Cantor model of DNA substitution and incorporating the gene conversion recombination model of \cite{Wiuf:2000} with exponential demographic growth, as simulation of this scenario is straightforward using existing tools (detailed below).  However, simulation of a large dataset is prohibitively slow so we used a random subsample of 100 isolates.  Coalescent theory suggests that such a sample size captures much of the information of the full sample \citep{Nordborg:2004}, and simulation based checks on informativeness are detailed in the supplementary material.  The selected isolates were confirmed to be \emph{C.~jejuni} using the PubMLST database and through a phylogeny analysis of these isolates and a representative \emph{C.~coli} sequence.  Three models were considered, with equal prior weights: \emph{Model 1} no growth; \emph{Model 2} growth for 50 years (since expansion of the New Zealand poultry industry); \emph{Model 3} growth for 170 years (since introduction of European livestock, primarily from Australia and the UK).

Each model has three biological parameters: a recombination rate, mean track length (i.e.~length of recombining DNA segment) and mutation rate.  Models 2 and 3 also have a growth parameter.  To aid interpretability we parameterised this as the relative increase in the effective population size during the period of growth.  Prior information on parameters is summarised in Table \ref{tab:prior}.  Mutation and recombination rates are given per kilobase per $2 N_e g$ years, where $N_e$ is the effective population size and $g$ the generation length in years.  \cite{Wilson:2009} estimated the mean time to coalescence, $N_e g$, at $218$ with an interval estimate of $[155,288]$.  To simplify our model, we fix $N_e g= 218$.  We expect that variations of $N_e g$ within the quoted interval will not affect the detection of growth.  Mean recombination length is in kilobase units.  The relative growth parameter is unitless as it is a ratio of effective population sizes.

Growth priors are based on demographics of the principal host; poultry for model 2 and sheep/cattle for model 3.  Rough estimates of host growth rates are used, based on the data of \cite{Binney:2013}, with variance increased to account for uncertainty of the link between bacterial and host demographics.  Biological parameter priors are based on analysis of other \emph{C.~jejuni} data in \cite{Wilson:2009}.  This assumed a no growth model, so these priors may not be appropriate for models 2 and 3.  Sensitivity analysis detailed in the supplementary material also considers a much less informative biological prior.

\begin{table}[htp] \scriptsize
\begin{center}
\begin{tabular}{*{7}{c}}
&&&&& \multicolumn{2}{c}{Log normal} \\
Parameter & Units & Model & Point estimate & 95\% CI & Mean & Sd \\
\hline
Mutation rate & $kb^{-1} (2N_e g)^{-1}$ & All & 13.7 & $[8.1,23.2]$ & 2.62 & 0.27 \\ 
Recombination rate & $kb^{-1} (2N_e g)^{-1}$  & All & 1.31 & $[0.03, 51.5]$ & 0.27 & 1.87 \\
Mean track length & $kb$ & All & 4.52 & $[0.1, 209.9]$ & 1.51 & 1.96 \\
Relative growth & - & 2 & 4.06 & $[1.5,10.8]$ & 1.40 & 0.50 \\
Relative growth & - & 3 & 33.1 & $[2.9, 383.8]$ & 3.50 & 1.25
\end{tabular}
\caption{Details of the parameter priors used in Section \ref{sec:appl}.  Priors are assumed to be the product of a log normal prior for each individual parameter.  The point estimates are geometric means.  The recombination length prior was truncated below 1 base pair, and the recombination rate above $25 kb^{-1} (2N_e g)^{-1}$ to avoid excessively slow simulations (All estimated posteriors for recombination rate were well below this - see Figure 2 of supplementary material.)}
\label{tab:prior}
\end{center}
\end{table}

\subsection{Methods}

Data sets were simulated using ms \citep{Hudson:2002} and seq-gen \citep{Rambaut:1997}.  Genetic summaries required were calculated using R \citep{R:2012}, which  was also used to code the inference algorithms.

We implemented semi-automatic ABC (Algorithm \ref{alg:semiauto2}) as follows.  First a pilot analysis was performed using the ABC SMC algorithm of \cite{Toni:2010} (detailed in the supplementary material) with 1000 particles.  This targeted log-transformed parameters, as on the original scale the target distribution is roughly log-normal and hard for the algorithm to explore.  The summary statistics were a set of 15 genetic summaries (these, and other summaries used below, are listed in the supplementary material).   The distance function was Equation \eqref{eq:scaled dist}, Euclidean distance between normalised summary statistics, with standard deviations estimated from 100 datasets simulated from the prior predictive distribution.  These simulations were also used to choose an initial ABC threshold: the median of the distances between these datasets and the observations.  In following SMC iterations, the threshold was the median of distances for accepted particles in the preceding step.  The algorithm terminated after the iteration which reached $2 \times 10^4$ simulated data sets.

To fit summary statistics, $2 \times 10^4$ datasets were simulated using the training regions.  Model choice summaries were fitted as described in Section \ref{sec:fit} and summaries for parameter inference by linear regression (detailed shortly).  For all regressions the vector of covariates $f(\cdot)$ consisted of 3 cubic B-spline bases for each of 125 genetic summaries, giving a total of 375 covariates, and a constant term.  Spline transformations were included to capture non-linear effects.  Due to the large number of covariates, $L_1$ penalised versions of logistic and linear regression were used, using the `glmnet' R package \citep{Friedman:2010} with the tuning parameter chosen by cross-validation.  Cross-validation estimates of fitting error were used to investigate which genetic summaries were most informative and to validate many of our modelling and tuning choices (details in supplementary material).

Exploratory analysis showed that for each parameter a single estimator could perform reasonably well under all models (details in supplementary material).  To keep $\dim(S)$ small, our $S$ is the concatenation of such estimators with model choice statistics.
A single hypercube training region was used for all models to prevent behaviour of a particular model being overrepresented in any region of parameter space.  This training region was the product of the parameter ranges from the entire pilot output, regardless of model.  The regression responses were log-transformed parameters, supported by exploratory analysis of Box-Cox transformations.  The resulting predictors were exponentiated to use in $S$.   Regressions for biological parameters were fitted using the simulations from all models, while those for the demographic parameter used simulations from the growth models only.    

The final $S$ vector used in the main ABC analysis consisted of four parameter estimators and three statistics for model choice.  The analysis used the distance function \eqref{eq:scaled dist} with summary statistic standard deviations estimated from the training data.  The analysis used the same SMC ABC algorithm as the pilot run, again with 1000 particles and targeting log-transformed parameters.  The initial threshold was the median of distances to the observed data calculated from the training data, with subsequent thresholds chosen as in the pilot run.  The algorithm terminated after the iteration which reached $4 \times 10^4$ simulated data sets.

\subsection{Results}

Table \ref{tab:results} summarises the model choice results for the pilot and main analyses, including the effect of regression post-processing as in \cite{Beaumont:2008}.  They agree in putting the majority of the weight on model 1, the no growth model.  Effective sample sizes \citep{Liu:1996} show that Monte Carlo error is approximately equal to that of a moderately large independent sample.
The supplementary material details sensitivity analyses which vary the parameter priors and the subsample of isolates used as observations.  With the exception of some pilot analyses, the weight placed on model 1 remains in the range $80-100\%$.  ABC analyses of simulated datasets are also described in the supplementary material.  Although only a small number were possible due to the high computational cost, the results suggest that the analyses are capable of distinguishing the no-growth from the growth models, with the main analysis doing so more accurately.

\begin{table}[htp]
\begin{center}
\begin{tabular}{ccc|ccc}
Analysis & ESS & Post-processed? & Model 1 & Model 2 & Model 3 \\
\hline
\multirow{2}{*}{Pilot} & \multirow{2}{*}{348} & No & 0.86 & 0.11 & 0.04 \\
 && Yes & 1.00 & 0.00 & 0.00 \\
\multirow{2}{*}{Main} & \multirow{2}{*}{600} & No & 0.96 & 0.03 & 0.01 \\
 && Yes & 0.92 & 0.03 & 0.05
\end{tabular}
\caption{Estimated posterior probabilities and effective sample sizes 
from ABC analyses on \emph{Campylobacter} data.}
\label{tab:results}
\end{center}
\end{table}

Table \ref{tab:param_post} summarises the parameter inference results.  Marginal density plots are provided in the supplementary material.  The table includes results from applying the regression adjustment of \cite{Beaumont:2002} to model 1 output.  This was not applied to other models as there were too few accepted particles to expect it to be stable.  The most notable finding is the low estimate of recombination rate, discussed further in Section \ref{sec:campy discussion}.  Additionally, informative estimates are made for mutation rate and relative growth.  The latter concentrates on low values, providing further evidence against significant growth.  Sensitivity analyses detailed in the supplementary material support these qualitative findings, although the numerical values are less robust than those for model choice.


\begin{table}[htp] \scriptsize
\begin{center}
\begin{tabular}{cc|cccc}
&& Recombination rate & Mean track length & Mutation rate & Relative growth \\
&& $kb^{-1} (2N_e g)^{-1}$ & $kb$ & $kb^{-1} (2N_e g)^{-1}$ &  \\
\hline
\multirow{3}{*}{Prior} & Model 1 & 1.31 [0.03, 51.5] & 4.52 [0.1, 209.9] & 13.7 [8.1, 23.2] & \\
& Model 2 & 1.31 [0.03, 51.5] & 4.52 [0.1, 209.9] & 13.7 [8.1, 23.2] & 4.06 [1.5, 10.8] \\
& Model 3 & 1.31 [0.03, 51.5] & 4.52 [0.1, 209.9] & 13.7 [8.1, 23.2] & 33.1 [2.9, 383.8] \\
\hline
\multirow{4}{*}{Pilot} & Model 1 & 0.34 [0.02, 5.21] & 2.43 [0.06, 88.2] & 11.4 [7.63, 16.7] & \\
& Model 1 (adjusted) & 0.18 [0.02, 1.87] & 1.04 [0.05, 24.8] & 12.8 [10.2, 16.7] & \\
& Model 2 & 0.28 [0.02, 2.45] & 1.99 [0.09, 24.1] & 12.6 [8.76, 17.2] & 2.12 [1.07, 3.07] \\
& Model 3 & 0.17 [0.01, 0.78] & 1.58 [0.08, 94.9] & 12.2 [8.80, 15.1] & 4.81 [0.97, 19.0] \\
\hline
\multirow{4}{*}{Main} & Model 1 & 0.55 [0.02, 3.74] & 5.81 [0.17, 239.2] & 12.9 [10.1, 16.5] & \\
& Model 1 (adjusted) & 0.22 [0.02, 1.18] & 2.98 [0.22, 63.2] & 13.0 [10.6, 15.9] & \\
& Model 2 & 0.24 [0.01, 3.53] & 5.73 [0.52, 239] & 14.0 [11.6, 16.5] & 1.51 [0.85, 2.71] \\
& Model 3 & 0.34 [0.01, 3.37] & 3.08 [0.40, 128] & 12.6 [9.81, 16.4] & 1.12 [0.41, 2.44]
\end{tabular}
\caption{Parameter point estimates (geometric means) and 95\% credible intervals from prior and ABC analyses on \emph{Campylobacter} data.}
\label{tab:param_post}
\end{center}
\end{table}


The regression and ABC results were also used to find which genetic summaries were particularly informative, and to show that some aspects of the data fitted poorly under any model.  These results are given in the supplementary material, and can inform future modelling and analyses.


\section{Discussion} \label{sec:discussion}

\subsection{ABC Methodology}


It is often desirable to perform model choice and parameter inference using the same simulations.  Our methodology focuses on producing $S$ appropriate for model choice only.  Section \ref{sec:appl} contains an application-specific example of adding a small number of further summaries to $S$ which are informative for parameter inference.  General purpose methods to choose such low dimensional summaries would be useful.  However, often each model may require separate summaries, so that a choice of $S$ suitable for model choice and parameter inference would be high dimensional.  An alternative strategy is to develop ABC methods in which comparisons of simulated and observed data do not always use the same summaries.  A simple approach would be to perform separate rejection sampling analyses for model choice and for parameter inference under each model.  A possible alternative is an MCMC algorithm which moves between models using only summaries relevant to the model(s) involved in the current step.

There are numerous alternatives to logistic regression to fit summary statistics for model choice, such as linear discriminant analysis \citep{Estoup:2012} and a comparison of their performance within ABC may be interesting.
Other parts of our semi-automatic method could also be varied.
For example, our choice of $S$ is a vector of one-to-one transformations of Bayes factors, and other transformations may perform differently.
Also, other methods could produce a more accurate training region, such as fitting a flexible model to the pilot output.


For simplicity we have used relatively simple ABC algorithms.  However, much progress is being made in improving algorithmic efficiency, especially of ABC SMC \citep[e.g.][]{DelMoral:2011}.  Our work is complementary to this and it could be used with many such improved algorithms.  Indeed ABC SMC algorithms can also be modified to incorporate semi-automatic ABC.  For example, recall that in Section \ref{sec:examples} the training data were reused as the simulations needed for ABC rejection sampling.  As suggested by \cite{Barnes:2012}, in ABC SMC they could be similarly reused for the first SMC iteration.

\subsection{\emph{Campylobacter} application} \label{sec:campy discussion}

Our main finding is support for a model with no change in the effective population size of \emph{C.~jejuni}.  This is surprising over a period where its ecological niche has greatly increased.  Analysis in the supplementary material shows some features of the data are poorly fitted under all models, suggesting that more detailed demographic structure is necessary to fit the data well.  One potential modification is
subpopulation structure amongst the hosts which might reveal that only some support growing \emph{C.~jejuni} populations.

Our analysis also produced parameter estimates.  Those for mutation rate and mean length of recombination tract are comparable to those from other work.  The point estimates of recombination rate are somewhat smaller than those of \cite{Wilson:2009}, who performed a similar ABC analysis on a different dataset.  Furthermore our credible intervals are much narrower, and exclude the estimates of \cite{Fearnhead:2005}, \cite{Biggs:2011} and \cite{Yu:2012}, who find recombination and mutation rates to be of the same order of magnitude.  The discrepancy with \cite{Wilson:2009} is conceivably due to their use of a heavy tailed prior or ABC tuning differences such as choice of threshold.  The others suggest differences in the model or data used.  For example, as discussed by \cite{Yu:2012}, their analysis, and that of \cite{Biggs:2011}, is for closely related sequences, and may reveal a high level of recombination that is then removed by purifying selection.


\paragraph{Acknowledgements}
The authors acknowledge the Marsden Fund project 08-MAU-099 (Cows, starlings and \emph{Campylobacter} in New Zealand: unifying phylogeny, genealogy, and epidemiology to gain insight into pathogen evolution) for funding this project. This publication made use of the \emph{Campylobacter} Multi Locus Sequence Typing website (http://pubmlst.org/campylobacter/) developed by Keith Jolley and sited at the University of Oxford (Jolley and Maiden 2010, BMC Bioinformatics, 11:595). The development of this site has been funded by the Wellcome Trust.

\section*{Appendix: Proof of Theorem \ref{thm:suff}}

Bayes sufficiency of $S(x)$ for $\M$ is equivalent to the following being true for all $i$ and $p_M$, and almost any $x$,
\begin{equation}
  \Pr(\M_i|S(x)) = \Pr(\M_i|x). \label{eq:suff}
\end{equation}
For convenience we introduce $\bm{p}=(p_M(\M_i))_{1 \leq i \leq M}$ to represent the prior mass function.  Also, let $\bm{1}$ be a vector of $M$ components equal to $1$.

First assume $S$ is Bayes sufficient for $\M$.  Define $h_i(S(x), \bm{p}) = \Pr_{\bm{p}}(\M_i | S(x))$ (i.e.~the conditional probability under prior $\bm{p}$) and note $h_i(S(x), \bm{p}) = \Pr_{\bm{p}}(\M_i | x)$.  The required function is $g(S(x)) = (h_i(S(x), M^{-1} \bm{1}))_{1 \leq i \leq_{M-1}}$.

It remains to prove Bayes sufficiency for $\M$ given a function $g$ of the form described in the theorem.  Henceforth we consider only the case $\bm{p} = M^{-1} \bm{1}$, since in this case \eqref{eq:suff} is equivalent to $\Pr(x|\M_i) = k \Pr(S(x)|\M_i)$ for some constant $k$, and applying Bayes' theorem to this proves \eqref{eq:suff} for general $\bm{p}$.  It also suffices to show that \eqref{eq:suff} holds for all $i<M$; the case $i=M$ follows as probabilities sum to 1.  Fix some $i<M$ 
and define an indicator variable $Y = \mathbb{I}[\M=\M_i]$.  Then $T_i(x) = \Pr(\M_i|x) = E[Y | x]$ and $\Pr(\M_i|S(x)) = E[Y | S(x)]$.  To prove \eqref{eq:suff}, we will show that these conditional expectations are almost always equal.  Standard properties of conditional expectation give $E[Y | S(x)] = E[E\{Y|x\} | S(x)] = E[T_i(x) | S(x)]$.  Finally, $E[T_i(x) | S(x)] = E[g_i(S(x)) | S(x)] = g_i(S(x)) = T_i(x) = E[Y|x]$ for almost all $x$ as required, where $g_i(\cdot)$ represents the $i$th component of the $g(\cdot)$ function.

\bibliography{choice}

\begin{thebibliography}{}

\bibitem[Atkinson and Cameron, 1993]{Atkinson:1993}
Atkinson, I.~A. and Cameron, E.~K. (1993).
\newblock Human influence on the terrestrial biota and biotic communities of
  {New Zealand}.
\newblock {\em Trends in Ecology \& Evolution}, 8:447--451.

\bibitem[Barnes et~al., 2012]{Barnes:2012}
Barnes, C.~P., Filippi, S., and Stumpf, M. P.~H. (2012).
\newblock Contribution to the discussion of {Fearnhead} and {Prangle} (2012).
  {Constructing} summary statistics for approximate {Bayesian} computation:
  {Semi-automatic} approximate {Bayesian} computation.
\newblock {\em Journal of the Royal Statistical Society: Series B}, 74:453.

\bibitem[Beaumont, 2008]{Beaumont:2008}
Beaumont, M.~A. (2008).
\newblock Joint determination of topology, divergence time, and immigration in
  population trees.
\newblock In Renfrew, C., Matsumura, S., and Forster, P., editors, {\em
  Simulation, Genetics and Human Prehistory}, pages 134--154. McDonald
  Institute Monographs.

\bibitem[Beaumont et~al., 2002]{Beaumont:2002}
Beaumont, M.~A., Zhang, W., and Balding, D.~J. (2002).
\newblock Approximate {B}ayesian computation in population genetics.
\newblock {\em Genetics}, 162:2025--2035.

\bibitem[Biggs et~al., 2011]{Biggs:2011}
Biggs, P.~J., Fearnhead, P., Hotter, G., Mohan, V., Collins-Emerson, J., Kwan,
  E., Besser, T.~E., Cookson, A., Carter, P.~E., and French, N.~P. (2011).
\newblock Whole-genome comparison of two \emph{Campylobacter jejuni} isolates
  of the same sequence type reveals multiple loci of different ancestral
  lineage.
\newblock {\em PloS One}, 6(11):e27121.

\bibitem[Binney et~al., 2013]{Binney:2013}
Binney, B., Biggs, P.~J., Carter, P., Holland, B., and French, N.~P. (2013).
\newblock Historical livestock importation into {New Zealand}.
\newblock {\em New Zealand Veterinary Journal}.
\newblock (submitted).

\bibitem[Blum, 2010]{Blum:2010}
Blum, M. G.~B. (2010).
\newblock Approximate {B}ayesian computation: A nonparametric perspective.
\newblock {\em Journal of the American Statistical Association},
  105(491):1178--1187.

\bibitem[Blum and Fran\c{c}ois, 2010]{Blum/Francois:2010}
Blum, M. G.~B. and Fran\c{c}ois, O. (2010).
\newblock Non-linear regression models for approximate {B}ayesian computation.
\newblock {\em Statistics and Computing}, 20:63--73.

\bibitem[{Del Moral} et~al., 2012]{DelMoral:2011}
{Del Moral}, P., Doucet, A., and Jasra, A. (2012).
\newblock An adaptive sequential {Monte Carlo} method for approximate
  {B}ayesian computation.
\newblock {\em Statistics and Computing}, 22(5):1009--1020.

\bibitem[Didelot et~al., 2011]{Didelot:2011}
Didelot, X., Everitt, R.~G., Johansen, A.~M., and Lawson, D.~J. (2011).
\newblock Likelihood-free estimation of model evidence.
\newblock {\em Bayesian Analysis}, 6(1):49--76.

\bibitem[Dingle et~al., 2001]{Dingle:2001}
Dingle, K.~E., Colles, F.~M., Wareing, D. R.~A., Maiden, M. C.~J., Ure, M.
  C.~J., Maiden, R., Fox, A.~J., Bolton, F.~E., Bootsma, H.~J., Willems, R.~J.,
  Urwin, R., and Maiden, M.~C. (2001).
\newblock Multilocus sequence typing system for \emph{Campylobacter jejuni}.
\newblock {\em Journal of Clinical Microbiology}, 39:14--23.

\bibitem[Estoup et~al., 2012]{Estoup:2012}
Estoup, A., Lombaert, E., Marin, J.-M., Guillemaud, T., Pudlo, P., Robert,
  C.~P., and Cornuet, J. (2012).
\newblock Estimation of demo-genetic model probabilities with approximate
  {B}ayesian computation using linear discriminant analysis on summary
  statistics.
\newblock {\em Molecular Ecology Resources}, 12(5):846--855.

\bibitem[Fearnhead and Prangle, 2012]{Fearnhead:2012}
Fearnhead, P. and Prangle, D. (2012).
\newblock Constructing summary statistics for approximate {B}ayesian
  computation: Semi-automatic {ABC}.
\newblock {\em Journal of the Royal Statistical Society, Series B},
  74:419--474.

\bibitem[Fearnhead et~al., 2005]{Fearnhead:2005}
Fearnhead, P., Smith, N. G.~C., Barrigas, M., Fox, A., and French, N. (2005).
\newblock Analysis of recombination in \emph{Campylobacter jejuni} from {MLST}
  population data.
\newblock {\em J Mol Evol}, 61:333--340.

\bibitem[Friedman et~al., 2010]{Friedman:2010}
Friedman, J., Hastie, T., and Tibshirani, R. (2010).
\newblock Regularization paths for generalized linear models via coordinate
  descent.
\newblock {\em Journal of Statistical Software}, 33(1).

\bibitem[Grelaud et~al., 2009]{Grelaud:2009}
Grelaud, A., Robert, C., Marin, J.-M., Rodolphe, F., and Taly, J.~F. ({2009}).
\newblock {ABC} likelihood-free methods for model choice in {G}ibbs random
  fields.
\newblock {\em {Bayesian Analysis}}, {4}({2}):{317--336}.

\bibitem[Hudson, 2002]{Hudson:2002}
Hudson, R.~R. (2002).
\newblock Generating samples under a {W}right-{F}isher neutral model of genetic
  variation.
\newblock {\em Bioinformatics}, 18:337--338.

\bibitem[Humphrey et~al., 2007]{Humphrey:2007}
Humphrey, T., O'Brien, S., and Madsen, M. (2007).
\newblock Campylobacters as zoonotic pathogens: a food production perspective.
\newblock {\em Int J Food Microbiol.}, 117(3):237--57.

\bibitem[Kolmogorov, 1942]{Kolmogorov:1942}
Kolmogorov, A.~N. (1942).
\newblock Determination of centre of dispersion and measure of accuracy from a
  finite number of observations (in {R}ussian).
\newblock {\em Izv. Akad. Nauk, USSR Ser. Mat.}, 6:3–32.

\bibitem[Liu, 1996]{Liu:1996}
Liu, J.~S. (1996).
\newblock Metropolized independent sampling with comparisons to rejection
  sampling and importance sampling.
\newblock {\em Statistics and Computing}, 6:113--119.

\bibitem[Marin et~al., 2012]{Marin:2012}
Marin, J.-M., Pillai, N., Robert, C.~P., and Rousseau, J. (2012).
\newblock Relevant statistics for {B}ayesian model choice.
\newblock {\em Preprint}.
\newblock Available at http://www.arxiv.org/abs/1110.4700.

\bibitem[Mullner et~al., 2009]{Mullner:2009}
Mullner, P., Spencer, S. E.~F., Wilson, D.~J., Jones, G., Noble, A.~D.,
  Midwinter, A.~C., Collins-Emerson, J.~M., Carter, P., Hathaway, S., and
  French, N.~P. (2009).
\newblock Assigning the source of human campylobacteriosis in {New Zealand}: A
  comparative genetic and epidemiological approach.
\newblock {\em Infection, Genetics and Evolution}, 9(6):1311--1319.

\bibitem[Nordborg, 2004]{Nordborg:2004}
Nordborg, M. (2004).
\newblock Coalescent theory.
\newblock In {\em Handbook of statistical genetics}, volume~2. Wiley.

\bibitem[Prangle et~al., 2013]{Prangle:2013}
Prangle, D., Fearnhead, P., Cox, M.~P., Biggs, P.~J., and French, N.~P. (2013).
\newblock Supplementary material for {S}emi-automatic selection of summary
  statistics for {ABC} model choice.
\newblock Available at http://www.arxiv.org/abs/???

\bibitem[{R Core Team}, 2012]{R:2012}
{R Core Team} (2012).
\newblock {\em R: A Language and Environment for Statistical Computing}.
\newblock R Foundation for Statistical Computing, Vienna, Austria.
\newblock {ISBN} 3-900051-07-0.

\bibitem[Rambaut and Grassly, 1997]{Rambaut:1997}
Rambaut, A. and Grassly, N.~C. (1997).
\newblock {S}eq-{G}en: {A}n application for the {M}onte {C}arlo simulation of
  {DNA} sequence evolution along phylogenetic trees.
\newblock {\em Computer Applications in the Biosciences}, 13:235--238.

\bibitem[Rayner and MacGillivray, 2002]{Rayner:2002}
Rayner, G.~D. and MacGillivray, H.~L. ({2002}).
\newblock Numerical maximum likelihood estimation for the g-and-k and
  generalized g-and-h distributions.
\newblock {\em {Statistics and Computing}}, {12}({1}):{57--75}.

\bibitem[Robert, 1996]{Robert:1996}
Robert, C.~P. (1996).
\newblock Intrinsic losses.
\newblock {\em Theory and decision}, 40(2):191--214.

\bibitem[Robert et~al., 2011]{Robert:2011}
Robert, C.~P., Cornuet, J.~M., Marin, J.-M., and Pillai, N. (2011).
\newblock Lack of confidence in approximate {B}ayesian computation model
  choice.
\newblock {\em Proceedings of the National Academy of Sciences},
  108(37):15112--15117.

\bibitem[Savill et~al., 2003]{Savill:2003}
Savill, M., Hudson, A., Devane, M., Garrett, N., Gilpin, B., and Ball, A.
  (2003).
\newblock Elucidation of potential transmission routes of \emph{Campylobacter}
  in {New Zealand}.
\newblock {\em Water Science and Technology}, 47(3):31--38.

\bibitem[Sears et~al., 2011]{Sears:2011}
Sears, A., Baker, M.~G., Wilson, N., Marshall, J., Muellner, P., Campbell,
  D.~M., Lake, R.~J., and French, N.~P. (2011).
\newblock Marked campylobacteriosis decline after interventions aimed at
  poultry, {New Zealand}.
\newblock {\em Emerging infectious diseases}, 17(6):1007--1015.

\bibitem[Sj{\"o}din et~al., 2012]{Sjodin:2012}
Sj{\"o}din, P., Sj{\"o}strand, A.~E., Jakobsson, M., and Blum, M. G.~B. (2012).
\newblock Resequencing data provide no evidence for a human bottleneck in
  africa during the penultimate glacial period.
\newblock {\em Molecular Biology and Evolution}, 29(7):1851--1860.

\bibitem[Sousa et~al., 2012]{Sousa:2011}
Sousa, V.~C., Beaumont, M.~A., Fernandes, P., Coelho, M.~M., and Chikhi, L.
  (2012).
\newblock Population divergence with or without admixture: selecting models
  using an {ABC} approach.
\newblock {\em Heredity}, 108:521--530.

\bibitem[Toni and Stumpf, 2010]{Toni:2010}
Toni, T. and Stumpf, M. P.~H. (2010).
\newblock Simulation-based model selection for dynamical systems in systems and
  population biology.
\newblock {\em Bioinformatics}, 26(1):104--110.

\bibitem[Wilson et~al., 2009]{Wilson:2009}
Wilson, D.~J., Gabriel, E., Leatherbarrow, A. J.~H., Cheesbrough, J., Gee, S.,
  Bolton, E., Fox, A., Hart, C.~A., Diggle, P.~J., and Fearnhead, P. (2009).
\newblock Rapid evolution and the importance of recombination to the
  gastroenteric pathogen \emph{Campylobacter jejuni}.
\newblock {\em Molecular biology and evolution}, 26(2):385--397.

\bibitem[Wiuf and Hein, 2000]{Wiuf:2000}
Wiuf, C. and Hein, J. (2000).
\newblock The coalescent with gene conversion.
\newblock {\em Genetics}, 155:451--462.

\bibitem[Yu et~al., 2012]{Yu:2012}
Yu, S., Fearnhead, P., Holland, B.~R., Biggs, P., Maiden, M., and French, N.~P.
  (2012).
\newblock Estimating the relative roles of recombination and point mutation in
  the generation of single locus variants in \emph{Campylobacter jejuni} and
  \emph{Campylobacter coli}.
\newblock {\em Journal of Molecular Evolution}, 74(5-6):273--280.

\end{thebibliography}


\end{document}